\documentclass[useAMS,usenatbib]{mn2e}
\usepackage{blindtext}
\usepackage{graphicx}
\usepackage{amssymb}
\usepackage{multirow}

\title[Fundamental Constant Observational Bounds on the Variability of the QCD Scale]{Fundamental Constant Observational Bounds on the Variability of the QCD Scale}
\author[Rodger I. Thompson]{Rodger I. Thompson$^{1}$\thanks{E-mail:
rit@email.arizona.edu (RIT)}\\
$^{1}$Steward Observatory, University of Arizona, Tucson, AZ 85721, USA}

\begin{document}

\date{Accepted xxxx. Received xxxx; in original form xxxx}

\pagerange{\pageref{firstpage}--\pageref{lastpage}} \pubyear{2016}

\maketitle

\label{firstpage}

\begin{abstract}
Many physical theories beyond the Standard Model predict time variations of basic physics
parameters.  Direct measurement of the time variations of these parameters is very difficult or
impossible to achieve.  By contrast, measurements of fundamental constants are relatively
easy to achieve, both in the laboratory and by astronomical spectra of atoms and molecules
in the early universe.  In this work measurements of the proton to electron mass ratio $\mu$ and
the fine structure constant $\alpha$ are combined to place mildly model dependent limits on the fractional
variation of the Quantum Chromodynamic Scale and the sum of the fractional variations of
the Higgs Vacuum Expectation Value and the Yukawa couplings on time scales of more than
half the age of the universe.  The addition of another model parameter allows the fractional
variation of the Higgs VEV and the Yukawa couplings to be computed separately.  Limits on their variation
are found at the level of less than $5 \times 10^{-5}$ over the past seven gigayears.  A model
dependent relation between the expected fractional variation of  $\alpha$ relative to $\mu$
tightens the limits to $10^{-7}$ over the same time span.  Limits on the present day rate of
change of the constants and parameters are then calculated using slow roll quintessence. A primary result of this
work is that studies of the dimensionless fundamental constants such as $\alpha$ and $\mu$,
whose values depend on the values of the physics parameters, are excellent monitors of the
limits on the time variation of these parameters.
\end{abstract}


\maketitle

\section{Introduction} \label{s-intro} 
Even though confirmation of the Higgs Boson \citep{aad12, cha12} and the detection of 
gravitational waves \citep{abb16} provide support for the Standard Model of physics and
General Relativity there are significant efforts to move beyond these theories.  A common prediction of many 
of these efforts is a time variation of basic physics parameters such as the Quantum 
Chromodynamic Scale, $\Lambda_{QCD}$, the Higgs Vacuum Expectation Value, $\nu$ 
and the Yukawa couplings, $h$, eg. \citep{cam95, cal02,lan02,lan04,din03,cha07,coc07,den08,
uza11}.  Detection of a variation of these parameters  would be a sure sign of physics
beyond the Standard Model while confirmation of their stability is consistent with
the Standard Model.  The values of two dimensionless fundamental constants, the fine 
structure constant $\alpha$ and the proton to electron mass ratio $\mu$ are functions
of $\Lambda_{QCD}$,  $\nu$ and $h$ therefore any variation of the parameters
produces a variation of $\alpha$ and $\mu$.

Reports of a possible time variation of $\alpha$ \citep{web01} initiated investigations
of the dependence of $\alpha$ and $\mu$ on $\Lambda_{QCD}$, $\nu$ and $h$
\citep{cal02,lan02,lan04,din03}.  These studies mostly concentrated on the
dependence of $\alpha$ and $\mu$ on the leading term, $\Lambda_{QCD}$.  Later work
by \citet{coc07} included the dependence on $\nu$ and $h$ as well. This forms the basis
of the relations developed in section~\ref{s-ppfc} which is an extension of the analysis in
\citet{thm16}. Whereas the previous works were centered on predicting changes in
$\alpha$ and $\mu$ from changes in $\Lambda_{QCD}$, $\nu$ and $h$ this work
centers on the bounds on time variability of $\Lambda_{QCD}$, $\nu$ and $h$ from
constraints on the time variability of $\alpha$ and $\mu$.  These constraints come
from astronomical observations of the value of $\alpha$ and $\mu$ in the early
universe over time scales on the order of the age of the universe.  To date there 
have been no uncontested observations of a change in either $\mu$ or $\alpha$.

Section~\ref{s-obs} discusses the current astronomical observations of $\mu$ and $\alpha$ 
with a concentration on the observations that provide the tightest constraint on a variation
of those constants.   The relationship between the physical parameters, $\Lambda_{QCD}, \nu ,h$
and the fundamental constants $\mu$ and $\alpha$ is developed in section~\ref{s-ppfc}. Section~\ref{s-cqcd}
develops the observational constraints on $\Lambda_{QCD}$ and the combination $\nu +h$,
with model dependent constraints on $\nu$ and $h$ independently.  The model of \citet{coc07} 
is examined in section~\ref{s-mod} and the constraints on the fractional variation of $\Lambda_{QCD}$.
$\nu$ and $h$ for the model are calculated. Also a model dependent limit on the variation of 
$\alpha$ based on the observational limit on the variation of $\mu$ is considered.  The limits
on evolution of the parameters in thawing and freezing forms of a quintessence cosmology 
are calculated in section~\ref{s-quin}. Section~\ref{s-evol} discusses the constraints on the
evolution of the parameters from the observational and model dependent constraints on the constants.  
Limits on the present day rates of change of the parameters and constants are also calculated in this section. 
Section~\ref{s-fo} discusses the direction of possible future observations and the conclusions are given 
in Section~\ref{s-cons}.

\section{Observations}  \label{s-obs}
The values of the fundamental constants $\mu$ and $\alpha$ are measured via astronomical 
spectroscopic observations of molecular and atomic lines, generally in absorption.  The absorbing
systems can be at significant redshifts giving time bases that are large fractions of the age of the 
universe.   The measured fractional variation is then $(c_z-c_0)/c_0$ where $c_z$ is the observed
value of a constant $c$ at the redshift of the observation and $c_0$ is the present day value of $c$
measured in the laboratory.  Below, the observations for $\mu$ and $\alpha$ are discussed separately.  
In each case the most stringent limit on a time variation of the constant is utilized in this work.

\subsection{The proton to electron mass ratio $\mu$ observations} \label{ss-mu}
The wavelengths of atomic absorption lines are relatively insensitive to variations $\mu$ but
molecular lines are good monitors of $\mu$ \citep{thm75}.  Approximately twelve quasar spectra
show absorption lines of molecular hydrogen produced in cold gas clouds along the line of sight
to the quasar.  At redshifts beyond 2 the absorption lines are redshifted from the ultra-violet into
the visible wavelengths which are observable with large ground  based telescopes and spectrometers
such as UVES at the VLT and HIRES at Keck.  Each line has a unique shift that depends on the 
quantum numbers of the upper and lower states that separates the $\mu$ shift from a redshift.
Accuracies on $\Delta \mu / \mu$ in the range of a few times $10^{-6}$ have been achieved
with this method, eg.~\citet{kin11}.

The H$_2$ lines are due to the Lyman and Werner electronic transitions.  Changes in $\mu$
most strongly affect the rotational and vibrational energies therefore the large electronic energies
dilute the fractional change in wavelength.  Recently high precision radio observations of 
molecules in their ground electronic and vibrational states have produced order of magnitude
more stringent constraints on a time variation of $\mu$.  The methanol \citep{jan11,lev11} and 
ammonia~\citep{fla07}
molecules are very sensitive to $\mu$ variations.  Observations of methanol lines in PKS1830-211
at a redshift of 0.88582 by \citet{bag13} and \citet{kan15} have restricted $\Delta \mu / \mu$ to
$(-2.9 \pm 5.7)  \times 10^{-8}$ where the error is the combined statistical and systematic $1 \sigma$ 
error.  Concerns about common lines of sight have increased the error to $\pm 1.0 \times 10^{-7}$
which is the constraint on the variation of $\mu$ used in this work.  The redshift of
this observation is relatively low compared to the $\alpha$ observations, but the look back time is 
$57\%$ of the age of a flat universe with $\Omega_m =0.3$ and $H_0 = 70.$.  The bound 
is equivalent to a linear time evolution of less than $\pm 7.88 \times 10^{-18}$ per year as compared 
to current atomic  clock measurements of $\pm 1.1 \times 10^{-16}$ per year \citep{god14}.
See, however, section~\ref{ss-pdt} for more realistic evolution models.

\subsection{The fine structure constant $\alpha$ observations} \label{ss-al}
The $\alpha$ observations are primarily astronomical optical spectroscopy of atomic fine structure.  There are 
several hundred high resolution and several thousand lower resolution spectra of high redshift fine
structure transitions.  There are also reports of
both a temporal and a spatial variation of $\alpha$ \citep{web01,web11} based on several sets 
of fine structure multiplets.  More recent observations, \citep{mur16,mol13} however, have not verified
these reports.  In particular \citet{mur16} attribute the reports of a variation of $\alpha$ to known 
errors in wavelength calibration.  For the purposes of this work we take the \citet{mur16} results as 
the primary set of observations and conclude that there is no validated variation of $\alpha$.

The new limits on $\Delta \alpha / \alpha$ in \citet{mur16} derive from observations of zinc and
chromium absorption lines that have a high sensitivity to changes in $\alpha$ for 9 quasar absorption
systems. Three systems were observed with both HIRES and UVES for a total of 12 independent 
observations. The weighted mean of these observations give $\frac{\Delta \alpha}{\alpha} = 0.4 \pm
1.7 \times 10^{-6}$ at the $1\sigma$ level where the error is the rms of the statistical and systematic errors.
This is a significantly lower constraint than the reported variation by \citet{web11} of $\approx -6.4 \pm 1.2
\times 10^{-6}$ and consistent with no change in $\alpha$.  The average redshift of the 12 observations is 1.54
which is a look back time of 9.4 gigayears or roughly $70\%$ of the age of the universe. The bound is
equivalent to a linear time evolution of less than $\pm 1.8 \times 10^{-16}$ per year which is a factor
of 10 less restrictive than the current laboratory limit \citep{god14} of $\pm 2.1 \times 10^{-17}$ per year. 
These observations are chosen as the primary observation set since the majority of the observations (63 hours)
were new observations with tight wavelength control.  The remaining are archival observations (38 hours)
with both UVES and HIRES. 

\section{Relating the physical parameters to the fundamental constants} \label{s-ppfc}
Having established the astronomical observational constraints on the time variation of
$\mu$ and $\alpha$ the functional relationship between variations in the constants and
variations in $\Lambda_{QCD}$, $\nu$ and $h$ are established next.  Although a variation 
of the physical parameters is allowed in this study, the dependence of the constants on the 
parameters in the Standard Model is maintained.  Under this assumption any variation of a fundamental 
constant requires a variation of the physics parameters that determine its value and vice versa.  
The discussion here follows the discussions in \citet{coc07} and \citet{thm16}
with an additional treatment of model dependent constraints on the time variation of
$\Lambda_{QCD}$, $\nu$ and $h$ individually.

\subsection{The proton to electron mass ratio $\mu$ relations} \label{ss-dmu}
The proton to electron mass ratio $\mu$ is probably the most obvious example of a 
relationship between a fundamental constant and the physics parameters. The fractional 
variation of $\mu$ is
\begin{equation} \label{eq-dmu}
\frac{d\mu}{\mu} = \frac{dm_p}{m_p} -\frac{dm_e}{m_e}
\end{equation}
A variation of $\mu$ requires a variation of the parameters that determine the mass of the proton 
and the mass of the electron.  The electron, as an elementary particle, by definition depends on the 
Higgs VEV and the electron Yukawa coupling as $m_e = h_e\nu$ therefore 
\begin{equation} \label{eq-dme}
\frac{dm_e}{m_e}=\frac{dh_e}{h_e} + \frac{d\nu}{\nu}
\end{equation}
The proton is not an elementary particle and its mass is a function of $\Lambda_{QCD}$, $\nu$ 
and the Yukawa couplings of the up and down quarks in the form 
$(\Lambda_{QCD})^a (\nu h)^b$ where $h$ refers to the quark Yukawa couplings and
$a$ and $b$ are scalars of order unity.  Since the units of this combination must be a 
mass the sum of the powers $a$ and $b$ must equal one for the proper dimensionality 
unless some other dimensional quantity has not been taken into account, from \citet{coc07}
where $a=0.76$ and $b=0.24$. This leads to the expression
\begin{equation} \label{eq-dmp}
\frac{dm_p}{m_p} = a \frac{d\Lambda_{QCD}}{\Lambda_{QCD}} + b (\frac{dh}{h}+\frac{d\nu}{\nu})
\end{equation}

Combining (\ref{eq-dme}) and (\ref{eq-dmp}) gives the expression for the fractional variation 
of $\mu$ as in \citet{thm16}
\begin{equation} \label{eq-dmu}
\frac{d\mu}{\mu} = a \frac{d\Lambda_{QCD}}{\Lambda_{QCD}} + (b-1) (\frac{dh}{h}+\frac{d\nu}{\nu})
\end{equation}
which uses the common assumption that the fractional variation of all of the Yukawa couplings
are similar and represented by $\frac{dh}{h}$.  If the condition that $(a+b)=1$ is invoked the
expression simplifies to
\begin{equation} \label{eq-qhy}
\frac{d\mu}{\mu} = a[\frac{d\Lambda_{QCD}}{\Lambda_{QCD}} - (\frac{dh}{h}+\frac{d\nu}{\nu})]
\end{equation}

From (\ref{eq-qhy}) a limit on the fractional variation of $\mu$ also limits
a combination of the fractional variation of $\Lambda_{QCD}$, $\nu$ and
$h$.  An appeal to naturalness might yield a similar limit on the individual fractional variations
but without more information no formal limits can be established.  Since $\alpha$ also depends
on the same parameters the limits on its variation can provide additional information.

\subsection{The fine structure constant $\alpha$ relations} \label{ss-da}
In \citet{coc07} the relation between the fractional change in $\alpha$ and the fractional
change in the physical parameters is given by
\begin{equation} \label{eq-dal}
\frac{d\alpha}{\alpha} = R^{-1}[\frac{d\Lambda_{QCD}}{\Lambda_{QCD}} - \frac{2}{9}(\frac{dh}{h}+\frac{d\nu}{\nu})]
\end{equation}
where it is again assumed that the fractional changes of the Yukawa couplings are similar. In
(\ref{eq-dal}) $R$ is a model dependent scalar which \citet{coc07} assumes to be $36$
but has a range of values in the literature.  The factor of $2/9$ is also model dependent.

\subsubsection{Some determinations of $R$} \label{sss-R}
Various authors have used different models and assumptions to 
determine estimates for the value of $R$.  One example is \citet{din03}
where the variation in $\alpha$ is produced by temporal changes of
the GUT unification scale $M_U$. There $R$ is given by
\begin{equation} \label{eq-R}
R=\frac{2 \pi}{ 9 \alpha}\frac{\Delta b_3}{\frac{5}{3}\Delta b_1+\Delta b_2}
\end{equation}
where the $b_i$  are the beta function coefficients between a scale $Q < M_U$ 
and $M_U$. At the unification scale all of the beta functions are unified to $b_U$.
$\Delta b_i$ is defined as $\Delta b_i  \equiv b_U-b_i$. The gauge couplings 
$\alpha_i(Q)$ $(i=1,2,3)$ are then given by
\begin{equation} \label{eq-cou}
(\alpha_i(Q))^{-1} = (\alpha_U(M_U))^{-1}-\frac{b_i}{2\pi} ln(\frac{Q}{M_U})
\end{equation}
The GUT scale $M_U$ is allowed to change but $\alpha_U(M_{Pl})$ and
$M_{Pl}$ are held constant where $M_{Pl}$ is the Planck mass.

At the unification scale $R$ is given by
\begin{equation} \label{eq-bu}
R=\frac{2 \pi}{ 9 \alpha}\frac{b_U+3}{\frac{8}{3}b_U - 12}
\end{equation}
As $b_U$ becomes either positively or negatively large
the value of $R$ approaches 36, the value used in \citet{coc07}. 

\section{Constraining $\Lambda_{QCD}$} \label{s-cqcd}
\begin{figure}
  \vspace{30pt}
\scalebox{.7}{\includegraphics{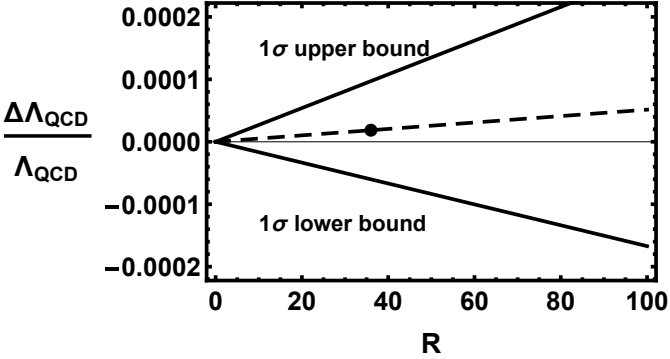}}
  \caption{The figure indicates the 1$\sigma$ variation of the limit on $\Delta \Lambda_{QCD} / \Lambda_{QCD}$ 
as a function of the model parameter $R$. The dashed line indicates the limit on $\Delta \Lambda_{QCD} / 
\Lambda_{QCD}$ if the measured value of the limits on $\Delta \alpha / \alpha$ and $\Delta \mu/\mu$ are used
rather than the limits.  The dot is at $R-36$ which is the example value.  Note that although it is not apparent at
the scale of the figure the limit on $\Delta \Lambda_{QCD} / \Lambda_{QCD}$ is not zero but rather the small
last term in (\ref{eq-lqcdm}) that does not depend on $R$.} 
\label{fig-qcdr}
\end{figure}

Equations (\ref{eq-dmu}) or (\ref{eq-qhy}) and (\ref{eq-dal}) provide two independent
equations in the two unknowns $\frac{\Delta \Lambda_{QCD}}{\Lambda_{QCD}}$ and 
the sum of the variations of the the Yukawa couplings and the Higgs VEV $(\frac{\Delta h}
{h} + \frac{\Delta \nu}{\nu})$.  These two equations are easily combined to eliminate one
of the unknowns.  Eliminating $(\frac{\Delta h}{h} + \frac{\Delta \nu}{\nu})$ yields
\begin{equation} \label{eq-dqcdn}
\frac{d \Lambda_{QCD}}{\Lambda_{QCD}} = \frac{d \alpha}{\alpha}\frac{(b-1)R}{[(b-1)-\frac{2}{9}a]}
+ \frac{d \mu}{\mu}\frac{2}{9[(b-1)-\frac{2}{9}a]}
\end{equation}
which is a function of the model parameters $R$, $a$, and $b$.  (\ref{eq-dqcdn}) simplifies if the
condition $(a+b)=1$ is invoked.
\begin{equation} \label{eq-dqcdab}
\frac{d \Lambda_{QCD}}{\Lambda_{QCD}} =\frac{9 R}{7}\frac{d \alpha}{\alpha} - \frac{2}{7 a}\frac{d \mu}{\mu}
\end{equation}
A similar exercise to eliminate $\frac{d \Lambda_{QCD}}{\Lambda_{QCD}}$ gives a constraint on
$(\frac{dh}{h}+\frac{d\nu}{\nu})$ of
\begin{equation} \label{eq-hvlim}
(\frac{dh}{h}+\frac{d\nu}{\nu}) = (\frac{9}{7})[R\frac{d \alpha}{\alpha}-\frac{1}{a}\frac{d \mu}{\mu}]
\end{equation}
Note that the leading terms on the right hand side of (\ref{eq-dqcdab}) and (\ref{eq-hvlim}) are identical.
In the following the 1$\sigma$ errors on the variation of $\mu$ $(\pm 10^{-7})$ and $\alpha$ $(\pm 1.7 \times 
10^{-6})$ quoted in sections~\ref{ss-mu}
and~\ref{ss-al} are used to compute the limits on the variation of the physics parameters.  It is assumed that
the errors are plus or minus about zero and not offset by the measured values.  Formally the errors are bounds
at the lowest redshift, 0.8858 for $\Delta \mu / \mu$.  At this redshift the error on the variation of $\alpha$ is
probably lower than the value measured at the average redshift of 1.54 assuming a monotonic evolution in
which the deviation of $\alpha$ from the present day value would be lower at the smaller redshift of the $\mu$
constraint.  Rather than attempting to estimate the reduction the error quoted at the higher redshift is retained
in the error budget.  With these values the constraints on the fractional variation of the parameters imposed by 
the limits on the fractional variation of the constants are given by
\begin{equation} \label{eq-lqcdm}
\frac{d\Lambda_{QCD}}{\Lambda_{QCD}} \leq \pm (1.7 \times 10^{-6})\frac{9 R}{7} \pm (1.0 \times 10^{-7})\frac{2}{7 a}]
\end{equation}
and
\begin{equation} \label{eq-lhvm}
(\frac{dh}{h}+\frac{d\nu}{\nu}) \leq \frac{9}{7} [\pm R(1.7 \times 10^{-6}) \pm (1.0 \times 10^{-7})\frac{1}{a}]
\end{equation}
Figure~\ref{fig-qcdr} shows the variation of the limit on $\Delta \Lambda_{QCD} / \Lambda_{QCD}$ as a function 
of $R$.

The $\pm$ in front of both terms in (\ref{eq-lqcdm}) and (\ref{eq-lhvm}) recognize that the error terms can be 
either negative or positive.  The total error is taken as the two terms in quadrature.  For values of the model 
parameter $R$ greater than unity the first term in 
(\ref{eq-lqcdm}) and (\ref{eq-lhvm}) dominates the constraint due to the much tighter constraint on a variation 
of $\mu$ than for $\alpha$.   Establishing stricter observational constraints on $\Delta \alpha / \alpha$ therefore yields 
the most improvement of the time variation constraints on $\Delta \Lambda_{QCD} / \Lambda_{QCD}$ and 
$(\Delta \nu / \nu + \Delta h /h)$. Due to the dominant and identical first terms both parameters have the same limit.  
(\ref{eq-lhvm}) is strictly a limit on the sum of $\Delta h / h$ and $\Delta\nu / \nu$.  An appeal to naturalness 
could say that the limit applies to both quantities individually, but, as shown in \S~\ref{ss-hvml}, the expected 
variation of $\nu$ is on the order of two magnitudes greater than the expected variation of the Yukawa couplings 
$h$.

\subsection{Model dependent limits on $h$ and $\nu$ individually} \label{ss-hvml}
The standard model establishes a relationship between the Higgs VEV and the Yukawa
couplings.  As an example \citet{coc07} gives a relationship between the Higgs VEV $\nu$ 
and the Yukawa coupling for the top quark $h_t$ as
\begin{equation} \label{eq-hnu}
\nu = M_{Pl}exp(-\frac{8 \pi^2 c}{h_t^2})
\end{equation}
where $M_{Pl}$ is the Planck mass and $c$ is a constant of order one.  The variation of $\nu$ and
$h_t$ are then coupled by 
\begin{equation} \label{eq-dhnu}
\frac{d \nu}{\nu}=16\pi^2c\frac{d h}{h^3} = \frac{158c}{h^2}\frac{d h}{h}\approx 160\frac{d h}{h}
\end{equation}
where the last term assumes that $c$ and $h$ are of order unity \citet{coc07} and we have again
assumed that $\frac{d h_t}{h_t}=\frac{d h}{h}$.  In this case the time variation of the Higgs VEV
is two orders of magnitude greater than the variation of the Yukawa couplings.  Since the factor multiplying 
$\frac{\Delta h}{h}$ is model dependent (\ref{eq-dhnu}) is often written simply as
\begin{equation} \label{eq-s}
\frac{d \nu}{\nu} = S\frac{d h}{h}
\end{equation}
Using (\ref{eq-s}) in (\ref{eq-hvlim}) yields
\begin{equation} \label{eq-dnu}
\frac{d \nu}{\nu} = \frac{9}{7}\frac{S}{(1+S)}[R \frac{\pm d \alpha}{\alpha} -  \frac{1}{a}\frac{\pm d \mu}{\mu}]
\end{equation}
Similarly the limit on $d h/h$ is
\begin{equation} \label{eq-h}
\frac{d h}{h} =  \pm\frac{9}{7}\frac{1}{(1+S)}[R \frac{\pm d \alpha}{\alpha} - \frac{1}{a}\frac{\pm d \mu}{\mu}]
\end{equation}
Note that these equations are different from equation (36) in \citet{coc07} which was obtained by holding 
$\alpha$ constant rather allowing $\alpha$ to vary and using the observational constraints on both $\mu$ and 
$\alpha$. 

\section{An Example Model} \label{s-mod}
The constraints developed in section~\ref{ss-hvml} are functions of the coefficients $R$, $S$, $a$ and $b$ 
which are set by the specific unification model employed.  The model of \citet{coc07} is an example where
the coefficients are $a = 0.76$, $b = 0.24$, $R=36$ and $S=160$.  For this set of coefficients the first term
or $\alpha$ term of the constraints dominates giving $1 \sigma$ constraints on the parameters as 
\begin{equation} \label{eq-qcdc}
\Delta \Lambda_{QCD} /  \Lambda_{QCD} \leq \pm 7.9 \times 10^{-5}
\end{equation}
\begin{equation}
\Delta \nu / \nu \leq \pm 7.9 \times 10^{-5}
\end{equation}
\begin{equation}
\Delta h / h \leq \pm 4.9 \times 10^{-7}
\end{equation} 
The  look back time for the constraints is the average look back time of the $\alpha$ observations at a
redshift of 1.54 equal to 9.4 gigayears or roughly $70\%$ of the age of the universe.

\subsection{Constraints with a Model Dependent $\Delta \alpha/\alpha$} \label{ss-ma}
The constraints used in section~\ref{s-mod} are the observational constraints on the time variation
of $\mu$ and $\alpha$.  These constraints, however, are not consistent with the example model.
The constraint on a variation of $\mu$ is more than a factor of 10 below the constraint on $\alpha$
whereas the model predicts that $\frac{\Delta \alpha}{\alpha} \approx \frac{1}{a R}\frac{\Delta \mu}{\mu}$
which makes the common assumption that the QCD scale dominates the evolution of both $\alpha$ and $\mu$ 
and that the $a$ coeficient in eqn.~\ref{eq-qhy} is of order unity.  Using $R=36$ from the previous model
yields a much lower limit on a variation of $\alpha$ such that $\frac{\Delta \alpha}{\alpha} \leq \pm
3.7 \times 10^{-9}$. The new model constraint is more than 450 times more restrictive than the
observational constraint.  Placing the new constraint into (\ref{eq-dqcdab}) yields
\begin{equation} \label{eq-dqcdmod}
\frac{\Delta \Lambda_{QCD}}{\Lambda_{QCD}} \leq [\pm (3.7 \times 10^{-9})\frac{9 R}{7} \pm 10^{-7}\frac{2}{7 a}]
\leq \pm 1.8 \times 10^{-7}
\end{equation}
Similar replacements in (\ref{eq-dnu}) and (\ref{eq-h}) yield constraints on the time variation of $\nu$ and $h$ of
\begin{equation} \label{eq-vmod}
\frac{d\nu}{\nu} \leq (\frac{9}{7})(\frac{S}{S+1})  (\pm R (3.7 \times 10^{-9})-\frac{\pm 10^{-7}}{a}]
\leq \pm 2.2 \times 10^{-7}
\end{equation}
\begin{equation} \label{eq-hmod}
\frac{d h}{h} \leq  (\frac{9}{7})(\frac{1}{S+1}) [(\pm R(3.7 \times 10^{-9})-\frac{\pm 10^{-7}}{a}]
\leq \pm 1.3 \times 10^{-9}
\end{equation}
Note that since the new limit on the variance of $\alpha$ is dependent on $R$ and $a$ eqns.~\ref{eq-dqcdmod}
-\ref{eq-hmod} are only valid for $R=36$ and $a=0.76$.

The new constraints on $\Delta \Lambda_{QCD} /  \Lambda_{QCD}$ are significantly more restrictive
than the observational constraints with a time scale of the look back time to the $\mu$ constraint at a 
redshift of 0.89 which is a little over $50\%$ of the age of the universe rather than the $70\%$ of the 
$\alpha$ observational constraint.  In the constraint on the variation of $\Lambda_{QCD}$ the $\alpha$
constraint is still the dominant term but the $\mu$ term is about a quarter of the $\alpha$ term and therefore
contributes to the quadrature sum.  In the constraints on $\Delta h /h$ and $\Delta \nu / \nu$ the $\alpha$
and $\mu$ constraints have roughly equal weight.  Since the constraint on $\alpha$ in this model is set
by the constraint on $\mu$, the limits can be improved by providing a tighter constraint on the time variation
of $\mu$.

\section{Quintessence Example of $\Lambda_{QCD}$ Evolution} \label{s-quin}
In the previous sections the source of the time variation of the physics parameters was not considered.
Here we examine a quintessence rolling scalar field as the source of dark energy. The coupling of the field
to the QCD scale and the Higgs VEV is the source of their variation as well.    In the 
following the evolution of $\Lambda_{QCD}$ and $\nu$ is considered by calculating the evolution of $\mu$
and $\alpha$ and consequently through (\ref{eq-dqcdab}) and (\ref{eq-dnu}) the evolution of 
$\Lambda_{QCD}$ and $\nu$.  By (~\ref{eq-s}) the evolution of $h$ is just $1/S$ times the evolution of
$\nu$.

Independently \citet{cal11} for $\alpha$ and \citet{thm12} for $\mu$ showed that for a rolling scalar field 
$\phi$ in a potential $V(\phi)$ with a quintessence dark energy equation of state
\begin{equation} \label{eq-w}
w =\frac{p_{\phi}}{\rho_{\phi}} =\frac{\dot{\phi}^2 -2V(\phi)}{\dot{\phi}^2 +2V(\phi)}
\end{equation}
the variation of $\mu$ or the fine structure constant $\alpha$ is  given by
\begin{equation} \label{eq-xint}
\frac{\Delta x}{x} =\zeta_{x}\int^{a(z)}_{1}\sqrt{3\Omega_{\phi}(a)(w_{de}(a)+1)}a^{-1}da
\end{equation}
The integral is over the scale factor $a$ from its present day value of $1$ to its value at the epoch of the 
observation $a(z)$. $\Omega_{\phi}$ is the ratio of the dark energy density to the critical density.  
$\zeta_x$ $(x= \mu, \alpha)$ is the strength of the coupling between the scalar field and $\mu$ or $\alpha$.  
The variation in $\mu$ or $\alpha$ is determined by the trajectory of the cosmological parameter
$w(a)$ and the magnitude of the new physics parameter $\zeta_x$.  Equation (\ref{eq-dqcdab}) then determines
 the variation of $\Lambda_{QCD}$ as a function of the scale factor $a$.
 
Freezing ($w$ evolves toward -1) and thawing ($w$ evolves away from -1) quintessence models are
used to limit the allowed cosmological evolution of $\Lambda_{QCD}$.   The form of the equation
of state is given by \citet{dut11} as
\begin{equation} \label{eq-qw}
1 + w = \frac{1}{3} \lambda_0^2[\frac{1}{\sqrt{\Omega_{\phi}}}-(\frac{1}{\Omega_{\phi}}-1)
(\tanh^{-1}(\sqrt{\Omega_{\phi}}) + C)]^2
\end{equation}
with
\begin{equation} \label{eq-c}
C = \pm \frac{\sqrt{3(1+w_i)}\Omega_{\phi_i}}{\lambda_0}
\end{equation}
C is determined at an early epoch where $\Omega_{\phi} \ll 1$ in which case $w_i$ and 
$\Omega_{\phi_i}$ are given by eqn.~\ref{eq-c} which is the limit of eqn.~\ref{eq-qw} for small
$\Omega_{\phi}$.  Inserting eqn.~\ref{eq-qw} and~\ref{eq-omega} into eqn.~\ref{eq-c} shows 
that $C$ is within $0.2\%$ of the value of $C$ put in eqn.~\ref{eq-qw} for all scale factors less than
0.1.

The value of $\lambda_0$ is set by the slow roll condition $\lambda^2 = (\frac{1}{V}\frac{dV}{d \phi})^2$ 
where the subscript $0$ indicates its present value. Since it is assumed to be constant it is the
value at all times.  The dark energy density factor $\Omega_{\phi}$ is given by
\begin{equation}\label{eq-omega}
\Omega_{\phi} = [1+(\Omega_{\phi_0}^{-1} - 1)a^{-3}]^{-1}
\end{equation}
where $\Omega_{\phi_0}$ is the current value of $\Omega_{\phi}$. Thawing solutions are given by the special 
case where $C=0$, $w_i = -1$.  A freezing case with $C=-1$ is used 
for comparison. A full discussion of these models is given by \citet{thm12}.  Note that the evolution
of $\alpha$ in (\ref{eq-xint}) is exactly similar to $\mu$ except for the value of the coupling constant 
$\zeta_{\mu,\alpha}$ which simply scales the evolution.  The evolution of the fractional changes in $\mu$ and $\alpha$
are found by numerical integration of (\ref{eq-xint}) with Mathematica using the functional forms of
$(1+w)$ and $\Omega_{\phi}$ given in (\ref{eq-qw}) and (\ref{eq-omega}).  Normally there is a
correction term in (\ref{eq-omega}) of the form $exp(3\int^{a}_{1}\frac{(1+w(x)}{3}dx)$ but direct
comparison of the cases considered here with and without the correction showed that the differences
were negligible  therefore the correction is ignored.  In particular the current value of $w+1$ for the 
thawing case is -0.0015, indistinguishable from 0.

The variables $\lambda_0$ and $\zeta_{\mu,\alpha}$ both multiply the integral in (\ref{eq-xint}) and
therefore scale the magnitude of the fractional change in $\mu$ and $\alpha$ but do not alter the shape of
the trajectory. Both of these variables are considered constant in time and either one or both can be used to
scale the changes to fit the observational constraints.  In the discussion of the evolution $\Lambda_{QCD}$
given below in section~\ref{s-evol} $\lambda_0$ is fixed at 0.1, satisfying the slow roll condition on $\lambda^2$, 
and all of the scaling is done by varying the couplings $\zeta_{\mu,\alpha}$.  For a chosen value of $C$ the
couplings must be less than or equal to the coupling values that meet the constraints on $\Delta \mu / \mu$
and $\Delta \alpha / \alpha$ at the redshifts of the constraint.

\section{Limits on the Evolution of $\Lambda_{QCD}$} \label{s-evol}
The limits on the evolution of  $\Lambda_{QCD}$ are examined for two different cases. The first case uses the 
observational limits on $\frac{\Delta \mu}{\mu}$ and $\frac{\Delta \alpha}{\alpha}$ and the second case uses 
the model limit on $\frac{\Delta \alpha}{\alpha}$ in section~\ref{ss-ma} derived from the observed limit on 
$\frac{\Delta \mu}{\mu}$.  In each case
the couplings $\zeta_{\mu}$ and $\zeta_{\alpha}$ are adjusted to satisfy the observed or modeled $1\sigma$ 
limits on $\frac{\Delta \mu}{\mu}$ and $\frac{\Delta \alpha}{\alpha}$ for the freezing and thawing cosmologies at
the redshifts of the constraints.  The integral in (\ref{eq-xint}) is then numerically integrated to calculate the fractional
variations of $\mu$ and $\alpha$ at scale factors between 0.2 and 1 corresponding to redshifts between 4 and 0.
Finally (\ref{eq-dqcdab}) calculates the trajectory of $\Delta \Lambda_{QCD} / \Lambda_{QCD}$ as a function
of scale factor limited by the maximum allowed variation of $\mu$ and $\alpha$ at each scale factor. 
Table~\ref{tab-ob} shows the allowed variations of the constants and the appropriate coupling constants for each  
case.  The calculations all use the example model parameter values for $a$, $R$, and $S$ given in section~\ref{s-mod} 
along with the assumption that $(a+b) = 1$.

\begin{table}
\begin{tabular}{llccccc}
\hline
Example & Cos.  & $\Delta \alpha/\alpha$ & $\Delta \mu / \mu$ & $\zeta_{\alpha}$ & $\zeta_{\mu}$\\
\hline
\multirow{2}{10mm}{Obs. Limits} & Frz. & 1.7E-6  & 1.0E-7 & -1.3E-5 & -1.3E-6 \\ 
& Thw. & 1.7E-6 & 1.0E-7 & -6.3E-5  & -4.4E-6\\
\hline
\multirow{2}{10mm}{Mod. Limits} & Frz. & 3.7E-9  & 1.0E-7 & -2.7E-8 & -1.3E-6 \\ 
& Thw. & 3.7E-9 & 1.0E-7 & -1.4E-8  & -4.4E-6\\
\hline
\end{tabular}
\caption{The relevant parameters for each of the evolution examples.  The entries in the cosmology (Cos.) column
are Frz. for the freezing cosmology and Thw. for the thawing cosmology.  The limits on $\Delta \alpha / \alpha$ and
$\Delta \mu / \mu$ are at the observational redshifts of 1.54 and 0.885 respectively.}  \label{tab-ob}
\end{table}

\subsection{Observationally Constrained $\Lambda_{QCD}$ Evolution}
\begin{figure} 
  \vspace{30pt}
\scalebox{.7}{\includegraphics{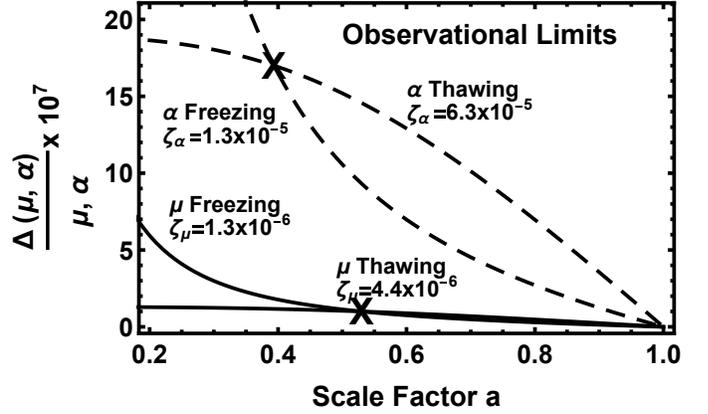}}
  \caption{The evolution of $10^7 \Delta \alpha / \alpha$ (dashed line) and $10^7 \Delta \mu /\mu$ (solid line)
for freezing and thawing cosmologies with $\zeta_{\mu}$ and $\zeta_{\alpha}$ set to observationally derivied 
values given in the Obs. Limits rows of Table~\ref{tab-ob}.  The X symbols where the trajectories cross mark the
$1\sigma$ limits on $\Delta \alpha / \alpha$ and $\Delta \mu / \mu$ which the trajectories must satisfy.}
 \label{fig-maob}
\end{figure}

\begin{figure}
  \vspace{30pt}
\scalebox{.7}{\includegraphics{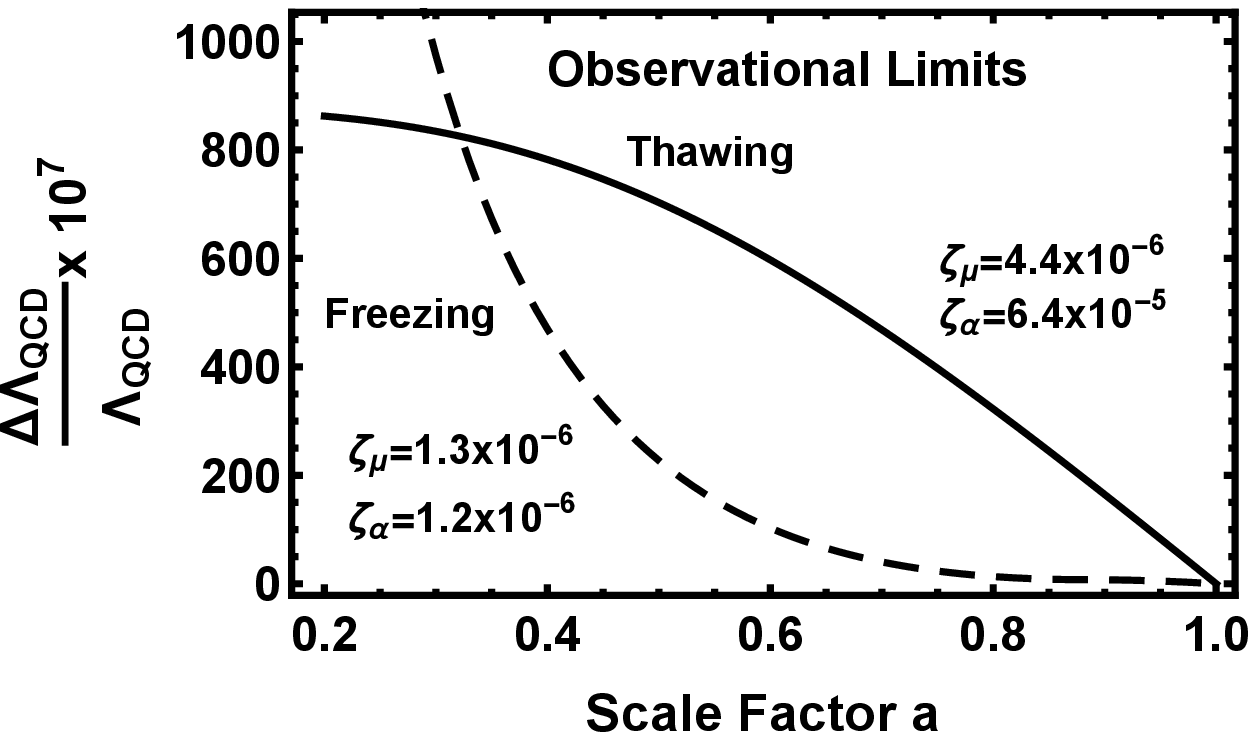}}
  \caption{The evolution of $10^7 \Delta \Lambda_{QCD}/\Lambda_{QCD}$ for the freezing (dashed line) and 
thawing (solid line) cosmologies with $\zeta_{\mu}$ and $\zeta_{\alpha}$ set to observationally derivied values 
given in the Obs. Limits rows of table~\ref{tab-ob}.}
 \label{fig-qob}
\end{figure}

Figure~\ref{fig-maob} shows the trajectories of $\Delta \alpha / \alpha$ and $\Delta \mu / \mu$ calculated with
$\mu$ and $\alpha$ coupling constants that satisfy the observational constraints for both the freezing and thawing
quintessence cosmologies.  For illustration the positive constraints are used but the negative equivalent is achieved
by flipping the sign of the coupling. It is clear that the significantly larger allowed error on $\Delta \alpha / \alpha$ along with
the larger coefficient of the $\alpha$ term in (\ref{eq-dqcdab}) make the $\alpha$ term dominant in determining 
$\Delta \Lambda_{QCD} / \Lambda_{QCD}$. Figure~\ref{fig-qob} shows the freezing and thawing evolution of 
$\Lambda_{QCD}$ calculated from the $\alpha$ and $\mu$ variations in fig.~\ref{fig-maob}.
This shows that the observationally constrained change in $\Lambda_{QCD}$ is limited to a factor of 
$\approx 5.0 \times 10^{-5}$ between a scale factor of 0.4 and 1.0 for both the freezing and thawing cases.  
At a scale factor 
of 0.2 the freezing cosmology change can be on the order of $2.4 \times 10^{-4}$ while the thawing change is limited 
to $\approx 8.5 \times 10^{-5}$.  

\subsection{Model Constrained $\Lambda_{QCD}$ Evolution} \label{ss-mqcd}
\begin{figure} 
  \vspace{30pt}
\scalebox{.7}{\includegraphics{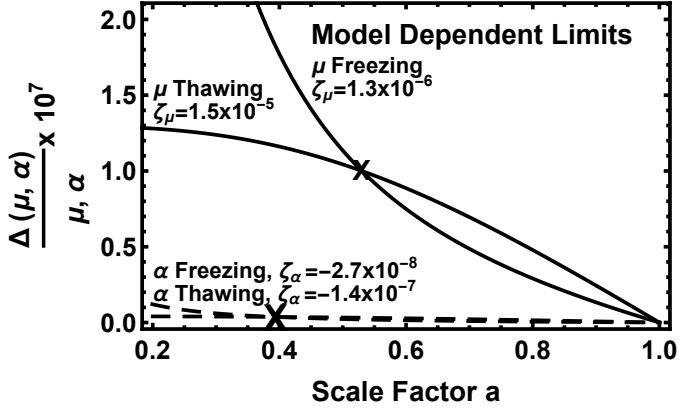}}
  \caption{The evolution of $10^7 \Delta \alpha / \alpha$ (dashed line) and $10^7 \Delta \mu /\mu$ (solid line)
for freezing and thawing cosmologies with $\zeta_{\mu}$ and $\zeta_{\alpha}$ set to observationally derived 
values given in the Mod. Limits rows of Table~\ref{tab-ob}.}
\label{fig-mamod}
\end{figure}

\begin{figure} 
  \vspace{30pt}
\scalebox{.7}{\includegraphics{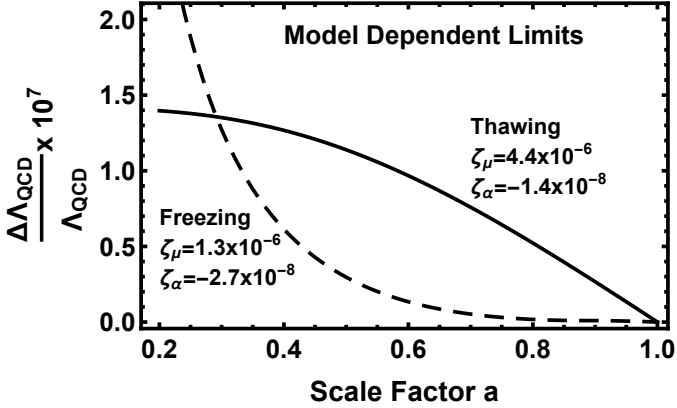}}
  \caption{The evolution of $10^7 \Lambda_{QCD}$ for the freezing (dashed line) and thawing (solid line) cosmologies
with $\zeta_{\mu}$ and $\zeta_{\alpha}$ set to observationally and modeled derived values given in the Mod. Limits rows
of table~\ref{tab-ob}.} 
\label{fig-qmod}
\end{figure}

Figure~\ref{fig-mamod} shows that the model constrained limit on $\Delta \alpha / \alpha$ is significantly reduced
by a factor of $1/27$ relative to the $\Delta \mu / \mu$ evolution. As shown in Fig.~\ref{fig-qmod} the model 
constrained case predicts very little $\Lambda_{QCD}$ evolution with changes in $\Lambda_{QCD}$ limited to 
$\approx \frac{\Delta \Lambda_{QCD}}{\Lambda_{QCD}} \le 3 \times 10^{-7}$ between a scale factor of 0.2 and 
the present scale factor of 1.  A scale factor of 0.2 is a redshift of 4.0 and a look back time of approximately 11.5 
gigayears.

\subsection{Limits on the current rate of change of $\Lambda_{QCD}$} \label{ss-pdt}
The integrand in (\ref{eq-xint}) is the derivative of $\mu$ or $\alpha$ with respect to the natural log of the scale
factor $ln(a)$.  Multiplying this by the Hubble Constant $\dot{a}/a$ converts the integrand to a time derivative.  The 
current rates of change of $\alpha$ and $\mu$ are then just the integrand at a scale factor of 1 multiplied by the 
current value of H, H$_0$. 

\begin{table}
\setlength{\tabcolsep}{3pt}
\begin{tabular}{llcccccc}
\hline
Limits & Cos.  & $\dot{\alpha}$ & $\dot{\mu}$ & $\dot{\Lambda}_{QCD}$ & $\dot{\nu}$ & $\dot{h}$\\
\hline
\multirow{2}{10mm}{Obs. Limits} & Frz. & 9.3E-17  & 8.9E-18 & 4.1E-15 & 4.1E-15 & 2.6E-17 \\ 
	& Thw. &2.7E-16 & 1.9E-17 & 1.2E-14  & 1.2E-14 & 7.6E-17\\
\hline
\multirow{2}{10mm}{Mod. Limits} & Frz. & 1.9E-19  & 8.9E-18 & -1.2E-17 & -2.4E-17 & -1.5E-19  \\ 
& Thw. & -5.9E-19 & 1.9E-17 & -9.7E-18  & -3.4E-17 & -2.1E-19 \\
\hline
Lab & - & $\pm$2.1E-17 & $\pm$1.1E-16 & -  & - & - \\
\hline
\end{tabular}
\caption{A table of the maximum allowed present day rates of change per year of the constants and 
parameters for the same examples and cosmologies as in Table~\ref{tab-ob}.  The laboratory limits 
on $\dot{\alpha}$ and $\dot{\mu}$ are from \citet{god14}.}  \label{tab-prat}
\end{table}

Table~\ref{tab-prat} gives the limits on the current rates of change per year of the constants and parameters for the
freezing and thawing cosmologies under the observational and model dependent limits.  The physics parameter
rates of change per year, $\dot{\Lambda}_{QCD}$, $\dot{\nu}$ and $\dot{h}$ are clearly too small to measure
on laboratory time scales.  The observational limit on $\dot{\alpha}$ is not as restrictive as the current laboratory
limits but the model dependent limit is two orders of magnitude more restrictive.  For $\dot{\mu}$  the observational
limits are an order of magnitude more restrictive than the laboratory results.  As expected the thawing cosmologies
allow a higher present day rate of change of the constants and the parameters than the freezing cosmology which
has a lower present day rate of change than the thawing cosmology.  This shows the dangers of using a linear rate
of change between the astronomically observed limits and the present day to predict the current rates of change.

\section{Future Observations} \label{s-fo}
Optical astronomical observations with high spectral resolution spectrometers on current and future large
telescopes may have the best chance of improving the limits the time evolution of the physical parameters.
This is particularly true at high redshift where the radio observations have yet to be accomplished.  The
radio observations at low redshift may have reached the limit where thermal and bulk motions of the 
absorbing gas could prevent more accurate constraints.  The highest existing or planned spectral resolution
optical spectrometer is the PEPSI instrument (R=300,000) recently installed on the Large Binocular Telescope.
The spectrometer is currently undergoing commissioning and has the potential for significant improvement
over previous observations.  The currently executing Large Program on the VLT \citep{mol13} also has great 
potential where the emphasis is on reducing the systematic errors in wavelength calibration through 
controlled observation programs and new calibration techniques as well as increasing the number of observations.
It is key in checking the claims of a variation in $\alpha$.

\section{Conclusions} \label{s-cons}
Measurements of the stability of the fundamental constants in the early universe provide an important tool
for the evaluation of new physics and cosmologies that predict time variation of the basic physics parameters that
determine the value of the constants.  In particular the combination of the observational limits on the time 
variation of the proton to electron mass ratio and the fine structure constant limit the fractional change of the
Quantum Chromodynamic Scale and the sum of the fractional changes of the Higgs Vacuum Expectation 
Value and the Yukawa couplings in terms of the model dependent parameters $R$ and $a$.  For the well
known model of \citet{coc07} where $R=36$ and $a=0.76$ the limit on the fractional change of the QCD
scale and the sum of the fractional changes of the Higgs VEV and the Yukawa couplings is $\leq 5 \times 
10^{-5}$ at the $1\sigma$ level for the last half of the age of the universe.  Further introduction of the 
model dependent parameter $S$ in the relation $d \nu /\nu = S dh/h$ provides individual limits on a 
variation of the Higgs VEV and the Yukawa couplings. 

The direct connection between the fundamental constants and the physics parameters provides a mechanism
to limit the time evolution of the parameters for rolling scalar field cosmologies.  The variation of $\mu$ and
$\alpha$ with time or scale factor is related to the evolution of the dark energy density and the the dark energy
equation of state \citep{cal11,thm12}.  The QCD Scale as a function of the scale factor $a$ in slow roll quintessence is 
examined as an example of such evolution.  A primary conclusion of the work is that the fundamental constants
play a central role in examining the validity of new physics and alternative cosmologies by defining the parameter
space of the theories that is consistent with observed limits on variability of the constants.

\label{lastpage}

\begin{thebibliography}{99}
\bibitem[\protect\citeauthoryear{Aad et al.}{2012}]{aad12} Aad, G. et al. 2012 Physics 
	Letters B, 716, 1
\bibitem[\protect\citeauthoryear{Abbott et al.}{2016}]{abb16} Abbott, B.P. et al. 2016, Phys. Rev. Lett.,
	116, 061102
\bibitem[\protect\citeauthoryear{Bagdonaite et al.}{2013}]{bag13} Bagdonaite, J.,Dapra, < Jansen, P.,  
	Bethlem, H.L., Ubachs, W., Henkel, C., \& Menten, K.M.  2013, Phys. Rev. Letters, 111, 231101
\bibitem[\protect\citeauthoryear{Calabrese et al.}{2011}]{cal11} Calabrese, E., Menegoni, E., M., Martins, C.J.A.P., 
	Melchiorri, A. \& Rocha, G. 2011, Phys., Rev. D, 84, 023518
\bibitem[\protect\citeauthoryear{Calmet and Fritzsch}{2002}]{cal02}	Calmet, X. and Fritzsch, H. 2002,
	Phys. Lett. B, 540, 173
\bibitem[\protect\citeauthoryear{Campbell and Olive}{1995}]{cam95} Campbell, B.A. and Olive, K.A. 1995,
	Phys. Lett. B, 345, 429
\bibitem[\protect\citeauthoryear{Chamoun et al.}{2007}]{cha07} Chamoun, N., Landou, S.J., 
	Mosquera, M.E. and Vucetich, H. 2007, J. Phys. G: Nucl. Part. Phys., 34, 163
\bibitem[\protect\citeauthoryear{Chatrchyan, S.  et al.}{2012}]{cha12} Chatrchyan, S. et al. 2012 Physics 
	Letters B, 716, 30
\bibitem[\protect\citeauthoryear{Coc et al.}{2007}]{coc07} Coc, A., Nunes, N.J., Olive, K.A., Uzan, J-P, \& Vangioni, E. 2007, Phys. Rev. D, 76, 023511
\bibitem[\protect\citeauthoryear{Dent}{2008}]{den08} Dent, T. 2008, Eur. Phys. J. Special 
	  Topics 163, 297–313 
\bibitem[\protect\citeauthoryear{Dine et al.}{2003}]{din03} Dine, M., Nir, Y., Raz, G. and Volansky, T.
	2003, Phys. Rev. D, 67, 015009
\bibitem[\protect\citeauthoryear{Dutta \& Scherrer}{2011}]{dut11} Dutta, S. \& Scherrer, R.J., Phys. Lett. B, 704, 265
\bibitem[\protect\citeauthoryear{Flambaum and Kozlov}{2007}]{fla07}	Flambaum, V.V. and Kozlov, M.G.
	2007, Phys. Rev. Let., 98, 240801
\bibitem[\protect\citeauthoryear{Godum et al.}{2014}]{god14} Godum, R.M., Nisbet-Jones, P.B.R., Jones, J.M.,
	King, S.A., Johnson, L.A.M., Margolis, H.S., Szymaniec, K., Lea, S.N., Bongs, K. \& Gill, P. 2014, Phys. Rev. Lett.,
	113, 210801 
\bibitem[\protect\citeauthoryear{Jansen et al.}{2011}]{jan11} Jansen, P., Xu, L.H., Kleiner, I., Ubachs, W.,
	and Bethlem, H. L., Phys. Rev. Let. 2011, 106, 100801
\bibitem[\protect\citeauthoryear{Kanekar et al.}{2015}]{kan15} Kanekar, N., Ubachs, W., Menten, K.M., 
	Bagdonaite, J., Brunthaler, A., Henkel, Muller, C.S., Bethlem, H.L. and Dapra, M. 2015, MNRAS 448, L104
\bibitem[\protect\citeauthoryear{King et al.}{2011}]{kin11} King, J. A., Webb, J. K., Murphy, M.,
        Ubachs, W, \& Webb, J. 2011, MNRAS, 417, 3010
\bibitem[\protect\citeauthoryear{Langacker et al.}{2002}]{lan02} Langacker, P., Segre, G. and
	Strassler, M.J. 2002, Phys. Lett. B, 528, 121 
\bibitem[\protect\citeauthoryear{Langacker}{2004}]{lan04} Langacker, P. 2004, Int. Jr. Mod. Phys. A,
	19, 157
\bibitem[\protect\citeauthoryear{Levshakov et al.}{2011}]{lev11} Levshakov, S.A., Kozlov, M.G. and
	Reimers, D. 2011, Ap.J., 738, 26
\bibitem[\protect\citeauthoryear{Molaro et al.}{2013}]{mol13} Molaro et al. 2013, A\&A, 555, A68
\bibitem[\protect\citeauthoryear{Murphy et al.}{2016}]{mur16} Murphy, M.T., Malec, A. and
	Prochaska, J.X. 2016, MNRAS 461, 2461
\bibitem[\protect\citeauthoryear{Thompson}{1975}]{thm75} Thompson, R.I. 1975, Astrophys. Lett., 15, 3
\bibitem[\protect\citeauthoryear{Thompson}{2012}]{thm12} Thompson, R.I., 2012, MNRAS Letters, 
	422, L67
\bibitem[\protect\citeauthoryear{Thompson}{2016}]{thm16} Thompson, R.I., 2016, Proceedings
	of the 14th Marcel Grossmann Conference arXiv:1602.03192v1 [astro-ph.CO]
\bibitem[\protect\citeauthoryear{Uzan}{2011}]{uza11} Uzan, J-P 2011, Living Rev. Relativity, 14, 2
\bibitem[\protect\citeauthoryear{Webb et al.}{2001}]{web01} Webb, J.K., King, Murphy, M.T., 
     Flambaum, V.V., Dzuba, V.A., Barrow, J.D., Churchill, C.W., Prochaska, J.X. \& Wolfe, A.M. 
	2001, PRL, 87, 091301-1-4
\bibitem[\protect\citeauthoryear{Webb et al.}{2011}]{web11} Webb, J.K., J.A., Murphy, M.T., 
     Flambaum, V.V., Carswell, R.F., \& Bainbridge, M.B. 2011, PRL, 107, 191101-1-5
\end{thebibliography}
\end{document}